\documentclass[letterpaper]{article} 
\usepackage[arxiv]{aaai24}  
\usepackage{times}  
\usepackage{helvet}  
\usepackage{courier}  
\usepackage[hyphens]{url}  
\usepackage{graphicx} 
\usepackage{natbib}  
\usepackage{caption} 
\usepackage{algorithm}
\usepackage{algorithmic}

\usepackage{newfloat}
\usepackage{listings}
\DeclareCaptionStyle{ruled}{labelfont=normalfont,labelsep=colon,strut=off} 
\floatstyle{ruled}
\newfloat{listing}{tb}{lst}{}
\floatname{listing}{Listing}
\title{Human-in-the-Loop AI for Cheating Ring Detection}
\author{
    Yong-Siang Shih,
    Manqian Liao,
    Ruidong Liu,
    Mirza Basim Baig
}
\affiliations{
    Duolingo, Inc. \\
    \{yongsiang,mancy,ruidong,basim\}@duolingo.com
}

\begin{document}

\maketitle

\begin{abstract}
Online exams have become popular in recent years due to their accessibility. However, some concerns have been raised about the security of the online exams, particularly in the context of professional cheating services aiding malicious test takers in passing exams, forming so-called "cheating rings". In this paper, we introduce a human-in-the-loop AI cheating ring detection system designed to detect and deter these cheating rings. We outline the underlying logic of this human-in-the-loop AI system, exploring its design principles tailored to achieve its objectives of detecting cheaters. Moreover, we illustrate the methodologies used to evaluate its performance and fairness, aiming to mitigate the unintended risks associated with the AI system. The design and development of the system adhere to Responsible AI (RAI) standards, ensuring that ethical considerations are integrated throughout the entire development process.

\end{abstract}

\section{Introduction}

Online exams, compared to test-center-based exams, offer greater accessibility for the test takers \cite{tan2021online}. However, contract cheating could post a challenge for maintaining the integrity and the validity of online assessment scores \cite{hill2021contract}. Contract cheating, also referred to as cheating ring, means that professional cheaters offer paid cheating services to help test takers cheat the exams. Such cheating behaviors, if uncaught, can greatly undermine the validity of the test scores. In this study, our aim is to develop a human-in-the-loop artificial intelligence (AI) system that automatically detects tests that could be involved in cheating rings. The suspicious tests will be escalated to human proctors for further scrutiny.

The proposed cheating ring detection system is developed within the context of a high-stakes online language assessment. The system's results have profound implications on test takers, as they inform whether a score, potentially utilizable for high-stakes purposes such as college admission, will be granted. As such, we, the developers, are obligated to ensure the system's accountability. This paper delves into the system's design, development, and evaluation, adhering to the RAI standards. We address key questions such as: 1) How does the system function and what is the nature of human-AI interaction within it? 2) How do we assess the system's performance and fairness? By addressing these questions, we illustrate the process of designing the system to meet the ethical principles and accountability required by Responsible AI (RAI) standards.

\begin{figure}[t]
\centering
\includegraphics[width=0.9\columnwidth]{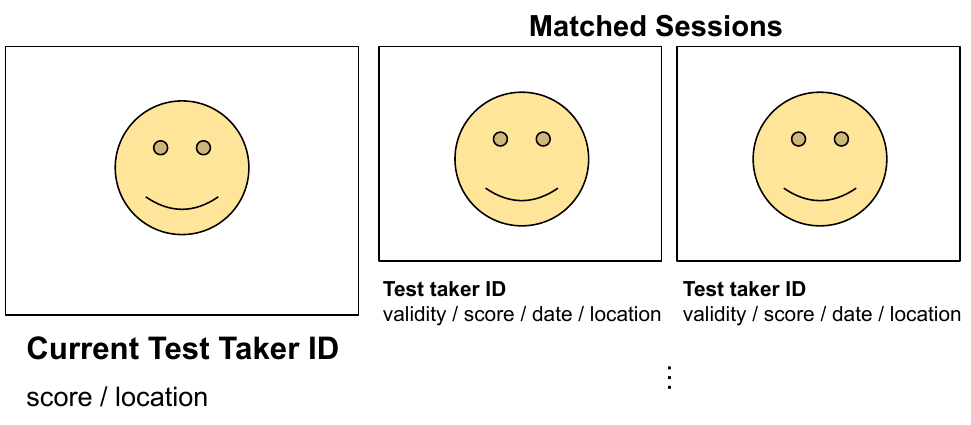} 
\caption{When proctoring the current test session, our cheating ring detection system may detect related test sessions to be shown to the proctors. The proctors can further investigate each test session by clicking on the test sessions. }
\label{fig.ui}
\end{figure}

\begin{table*}[t]
    \centering
    \begin{tabular}{| l | c c c | c c c c c c c  |}
    \hline
           & \multicolumn{3}{c|}{gender} & \multicolumn{7}{c|}{age} \\ \hline
    group & female & male & others & 0-14 & 15-20 & 21-25 & 26-30 & 31-35 & 36-40 & 41- \\ \hline
    ratio & 62.00\% & 62.36\% & 0.28\% & 4.42\% & 52.35\% & 37.96\% & 17.10\% & 10.18\% &5.63\% & 5.22\% \\ \hline
    TNR & 99.44\% & 99.45\% & 100\% & 100\% & 99.48\% & 99.28\% & 99.35\% & 99.87\% &100\% & 100\% \\ \hline
    \end{tabular}
    \caption{Performance across different groups for \emph{deep-keystroke+mouse} method. } \label{tbl.fair}
\end{table*}

\section{System Overview}

This section offers an overview of the cheating ring detection system. The system is intended to be integrated into the existing online examination platform. When a test taker takes a test, several data are collected including a recorded video, machine-specific information, network statistics, and details regarding keystrokes and mouse movements.

While contract cheating can manifest in numerous ways, this paper mainly focuses on the detection of services wherein professional cheaters, employed by the cheating ring service, seize control of the test taker's keyboard and mouse. The cheaters may either take the tests on behalf of the test takers, impersonating their identities, or merely control the periphery while the test takers pretend to take the test on their own. It operates under the assumption that these professional cheaters have impersonated multiple test takers. Consequently, we leverage the patterns of keystrokes and mouse movements to pinpoint test sessions that are likely to be impersonated by the same individual, serving as a method to identify potential cheating rings. The proposed cheating ring detection system consists of two components:

\subsubsection{Keystroke and Mouse Movement Pattern Analysis} For each test session, we compare its keystroke and mouse movements against those of the historical sessions. A test session is flagged as suspicious and potentially involved in a cheating ring if its keystroke and mouse movement patterns exhibit excessive similarity to those of another test taker.

\subsubsection{Proctoring Interface}
When a test is suspected of involvement in a cheating ring, the test, along with the historical tests exhibiting similar keystroke/mouse movement patterns, are presented to the proctors via the system interface as shown in Figure~\ref{fig.ui}. A camera image captured during each test and some important properties are displayed in a single page so that proctors can quickly compare the tests. When proctors click on each test, a detailed page would be opened for further investigation. If multiple historical tests exhibit patterns similar to the current test, the keystroke/mouse movement similarity index is used to rank the level of suspicion associated with each historical test.

\section{System Evaluation}

In this section, we describe our experiments to evaluate the keystroke and mouse movement pattern analysis methods. The dataset was sampled from the test sessions taken in the Duolingo English Test in 2022 and 2023, and is composed of around 127k test takers. We split the test takers into training, validation, and test splits with a 6:2:2 ratio. The positive pairs are constructed by sampling two test sessions from the same test taker while the negative pairs are constructed by sampling two test sessions from different test takers that are using the same type of keyboards and mice or are located in a similar region. There are roughly 6k positive pairs and 6k negative pairs in the validation and the test set, while the training pairs are randomly sampled during training.

The compared methods include a \emph{keystroke} baseline, which is t-test method based on the work of \citeauthor{young2019keystroke}~\shortcite{young2019keystroke}, while the other ones are neural networks using keystrokes \cite{young2019keystroke} and mouse \cite{zheng2011efficient} features, and were trained with a modified n-pair loss objective \cite{sohn2016improved} with L-2 distance.

Given a pair of test sessions, each method predicts a score indicating the similarity between the two sessions. We use the validation split to select a threshold for each method to make a prediction that achieves a FPR smaller or equal to 1\%, and then evaluate the models on the test set. The results are shown in Table~\ref{tbl.perf}. When using only keystorkes features, the deep model performs better on the AUROC metric compared to the baseline, but the FNR is worse on the FPR level that we choose. The best performance is achieved by the deep model using both keystrokes and mouse features.

\begin{table}
    \centering
    \begin{tabular}{| l c c c  |}
    \hline
    Method & AUROC & FPR & FNR \\ \hline
    keystroke & 95.69\% & 0.88\% & 20.20\% \\
    deep-keystroke & 97.96\% & 1.04\% & 27.61\% \\
    deep-mouse & 93.99\% & 1.06\% &  44.36\% \\
    deep-keystroke+mouse & \textbf{98.74}\% & \textbf{0.58}\% & \textbf{12.35}\% \\ \hline
    \end{tabular}
    \caption{Performance for different methods.} \label{tbl.perf}
\end{table}

We conducted studies on the fairness of the method. Here we adopt the notion of \emph{equality of opportunity} \cite{hardt2016equality}, and aim to equalize the true negative rate (TNR) across different demographic groups. We found that, under the same chosen cutoff, a similar level of TNR (around 98\%-99\%) are achieved across different groups, including ages, genders, and regions for the \emph{deep-keystorke-mouse} method. Some of the results are shown in Table~\ref{tbl.fair}. Note that a pair might involve two test takers from different demographic groups, and such a pair is counted in both groups, so the sum of ratios exceeds 100\%.

\section{Discussion}

In this paper, we introduced a system designed to detect cheating rings in online high-stakes assessments, aimed at protecting the security and integrity of the assessments.

There are several limitations in the study, which present opportunities for further research. First, our fairness assessment was primarily based on the TNR metric, which, while ensuring fair treatment of innocent test takers across different groups, does not guarantee uniform detection of cheaters across these groups. This calls for future exploration of a broader set of fairness metrics to more comprehensively evaluate the system's fairness. Second, our evaluation focused on the errors in the AI system without addressing the potential errors made by human proctors. A future area of research could involve investigating the fairness and biases in human proctoring decisions. 

Cheating detection in high-stakes assessments has significant implications for both test takers and score users. While the performance and fairness evaluation indicated that the AI signals in the proposed system are promising in detecting cheating ring, we emphasize that these signals should not be the sole determinant in cheating accusations of individuals. Instead, they should be integrated as crucial evidence, complementing other factors in the decision-making process. 

In practical terms, deploying the proposed system necessitates adherence to responsible AI standards, which include a commitment to protecting the privacy of test takers and preventing potential societal biases. In addition, it is necessary to continuously adapt the system to counter the evolving
cheating methods.

\bibliography{aaai24}

\end{document}